# Polymer Brushes and Grafted Polymers: AI/ML-Driven Synthesis, Simulation, and Characterization towards autonomous SDL

Rigoberto C. Advincula*[1,2] and Jihua Chen[1]

[1]Center for Nanophase Materials Sciences, Oak Ridge National Laboratory (ORNL)
1 Bethel Valley Road, Oak Ridge, TN 37830

[2]Department of Chemical and Biomolecular Engineering,
University of Tennessee at Knoxville
1512 Middle Dr, Knoxville, TN 37996

* To whom correspondence should be addressed: radvincu@utk.edu

This manuscript has been authored by UT-BatteIIe, LLC, under Contract No. DEAC05-00OR22725 with the U.S. Department of Energy. The United States Government and the publisher, by accepting the article for publication, acknowledges that the United States Government retains a nonexclusive, paid-up, irrevocable, world-wide license to publish or reproduce the published form of this manuscript, or allow others to do so, for United States Government purposes. DOE will provide public access to these results of federally sponsored research in accordance with the DOE Public Access Plan
(http://energy.gov/downloads/doe-public-access-plan).

# ABSTRACT

Polymer brushes and grafted polymers have attracted significant interest at the intersection of polymers, interfacial chemistry, colloidal science, and nanostructuring. The confinement of high-density grafted polymers and differences in swelling regimes govern the synthetic challenges and the interesting physics underlying their macromolecular dynamics. In this article, we focus on another intersection, artificial intelligence and machine learning (AI/ML), and how workflows will enhance the microstructure and composition of these systems. It will also accelerate potential applications through high-throughput experimentation (HTE) and data-driven intelligence, enabling scientific discovery and optimization. Applications in microfluidics, sensors, bioimplants, drug delivery, and related areas may yet offer more opportunities for ML-driven optimization. There is also interest in applying these studies with self-driving laboratories (SDLs) that can leverage autonomous systems for synthesis screening, characterization, and application evaluation.

# 1. Introduction of polymer brushes

Polymer science is undergoing a transformation to converge simulation, chemical informatics/data analytics, and artificial intelligence/machine learning (AI/ML), as well as high-throughput experiments, including synthesis and characterization.[1] The pillars of polymer science and engineering encompass many research areas at the intersection of chemistry, physics, and engineering. And the scope is not limited to solution properties or solid state. Polymer brushes, or grafted polymers, are an interfacial and surface phenomenon, like another state or subclass of polymer research, that will also undergo a specific convergence related to AI/ML ecosystems.

The confinement of polymers within specific surface-to-volume and interphase regimes is intriguing because it lies at the boundaries of interfacial phenomena and colloidal science.[2] Special interests in adhesion, tribology, particle handling, composites, etc., immediately give potential applications from translational research of their fundamental studies. Polymers can be grafted onto-, from-, and through- surfaces, using different chemical approaches.[3] This has been demonstrated through free-radical, living-free-radical, anionic, cationic, metathesis, click-chemistry, electron-transfer, photo-activation, polymerization, etc.[4] These methods, similar to those typically explored in solution, bulk, and emulsion polymerization, lead to the creation of ultrathin polymer layers on surfaces or at interfaces. What is intriguing is the ability to control both the density and the swelling regimes. High grafting density ($d$) with respect to the radius of gyration ($Rg$) and stretch regimes (conformation) resembles brushes – thus the term polymer brushes. (**Figure 1**) To minimize contact between polymer chains and maintain entropic freedom, the chains are stretched out. In addition, stimuli-responsive properties can be demonstrated based on grafting density, chain composition, or block composition, microstructure, and field response (temperature, light, chemical, etc.).[5,6] This can be applied to any substrates: flat surfaces, colloidal particles, nanoparticles, heterogeneous surfaces, microfluidic channels, polymers, as well as the interphase of two media: solid-gas, solid-liquid, liquid-gas, and liquid-liquid.[7] It should be noted that the term polymer brush applies to the dimensionality of the substrate and could also refer to grafting on a single polymer chain (1D), a flat substrate (2D), and a particle or heterogeneous surface (3D).[8]

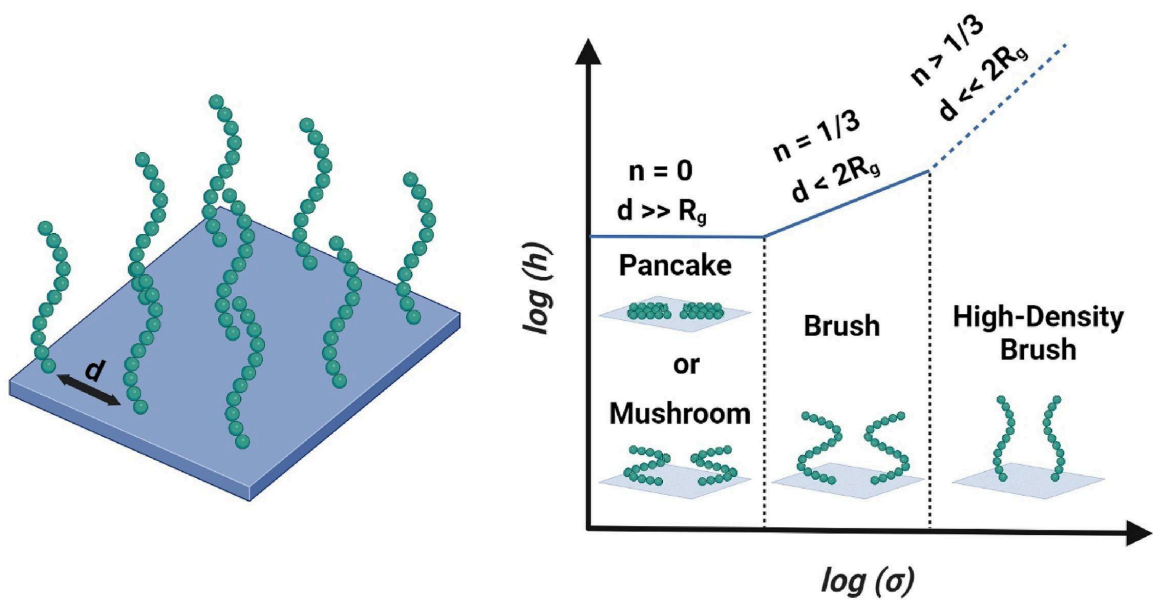

**Figure 1**. Schematic showing correlation between grafting density and conformation of grafted polymers (pancake, mushroom, brush, and high-density brush) in a good solvent [9] (Figure and caption from reference [9] with no change under a creative commons license BY-NC 4.0 [10])

In tribology, the polymer brush properties and the presence of a solvent variable affect load, electrostatic charges, and osmotic pressure, making it ideal for studies in lubrication and the characterization of friction coefficients.[11] In biological and biomedical applications, there are opportunities for non-fouling surfaces, biosensors, drug delivery, lab-on-a-chip, and lubrication.[12,13] In both lithographic and non-lithographic patterning, including nanopatterning, chemistries can be localized or incorporated into multi-step procedures, resulting in high contrast of surface and topological features.

The hypothesis-driven science and potential applications are well reported, but there are many variables, parameters, synthetic methods, compositions, microstructures, and even overlapping characterization methods.[14] The body of literature in the synthesis, chemistry, physics, simulation, patterning, and characterization is very large. The large language models (LLM)s for data mining and agentic AI tasks for their further research and development will take advantage of this body of literature.[15] Retrieval-augmented generation (RAG) tasks can be very specific and yield many answers, closer to the desired solution.[16]

Innovative characterization techniques applied to grafted systems and polymer brushes are intriguing.[17] Starting with ultrathin films and solid-state confinement, techniques such as FT-IR-ATR, XPS, AFM, SPR, ellipsometry, EIS, QCM, SIMS, X-ray reflectometry, and neutron reflectometry can be used.[18–23] With colloidal and nanoparticle substrates, light scattering, optical microscopy, fluorescence microscopy, TEM, SEM, X-ray, and neutron scattering, etc. can be applied.[24–26]

From thickness measurements, dielectric properties, polymerization kinetics, diffusion kinetics, ion and electron transport behavior, and tribology, it is of great interest to apply ex-situ, in-situ, real-time, or hyphenated experiments and measurements on these systems. Again, LLMs, RAGs, etc. will have a large body of literature and language models to distill the science. So, what is the interest in bringing more AI/ML workflows to this system?

One of the key limitations in any scientific discovery, optimization, or engineering-related study is the lack of training data and high-throughput experimentation (HTE) systems that will transform the field's ontology. By bringing an end-to-end AI/ML workflow to this field, it is possible to achieve the following: 1) Rapidly introduce novices to this field and accelerate the scientific skills build-up, 2) enable higher translational research from discoveries on their fundamental studies, 3) enable faster optimization protocols on their development for a particular application, and 4) enable a virtuous cycle of data-feedback-experimentation (DFE) to improve the whole system. While this can generally be said of any field for AI/ML augmentation, AI-assisted polymer brush R&D will accelerate many applications in coatings, formulations, healthcare, biomaterials, and other industries that rely on macromolecules at interfaces.

## 2. Synthetic routes, Interfaces, and Characterization

Most reported polymerization methods at interfaces are based on chain-addition growth mechanisms.[8] This is because initiators can be attached to the surface, enabling initiation, propagation, chain transfer, and termination to be controlled and monitored in thin films. Also, different block copolymer sequences are possible.[27] However, polymer brushes can be approached in general with three methods. Surface-initiated polymerization (SIP) is initiated primarily by surface initiators and is a grafting-from method. It affords high-density tethered polymers, which are also known as polymer brushes. With grafting methods, the population of monomers on surfaces for propagation results in lower-density brushes. Grafting-on strategies provide high versatility in synthetic mechanisms, including step-growth reactions, ring-opening metathesis, etc., since the polymer is pre-made before tethering. The grafting-on method is based on the physisorption or chemisorption of pre-synthesized polymers. The chain ends, or multi-regional functional groups, can adsorb onto the surface via a selective phase or a high reaction affinity for the corresponding polymer functional group. However, the enthalpic and entropic considerations do not favor a high-density brush. It can have many uses and be a cost-effective method for industrial scale-up, but it is not ideal for polymer brush studies. Lastly, the grafting-through strategy adds layers of a polymerizable group, monomer, or crosslinker, etc., that is not the initiator but participates in "stitching" the growing polymer chain or participates in chain transfer reactions.[28]

The grafting-from strategy, or SIP, began as a novelty because it allows the creation of high surface densities of initiators or a chain-transfer agent/initiator and has been applied to particle surfaces for some time.[29] It can be considered a complement to nanostructuring, patterning, and macromolecular control that originates at the surface or the substrate. This results in linear control over thickness and block copolymer layering, and it is the subject of many mechanistic studies of polymerization reactions at surfaces and interfaces, which have opened novel and reliable characterization methods via spectroscopy, optical methods, microscopy, dielectric methods, and reflectometry. For colloidal or nanoparticle substrates, the core-shell and hollow-shell or interfacial contrasts can be monitored using a range of methods, from light scattering to TEM.[30]

Mechanisms for SIP include free-radical, living-free-radical, anionic, cationic, metathesis, click chemistry, electron transfer, photoactivation, and polymerization. Their nuances are well documented for their specific advantages and versatility. The most widely used SIP methods are controlled radical polymerization (CRP) and living free radical polymerization (LFRP).[31] The most common methods are atom transfer radical polymerization (ATRP)[32] and reversible addition fragmentation and chain transfer RAFT polymerizations.[33] Many variations in terms of radical activation, redox activation, photo-activation, electron-transfer, metal-activation methods, etc., have been employed.[34–36] Substrates and surfaces include flat or rough surfaces, colloidal particles, nanoparticles, heterogeneous surfaces, microfluidic channels, polymers, as well as the interphase of two media.

Our early contributions are in the area of adopting anionic polymerization by grafting from-surfaces. An early review on the grafting of polymer brushes by anionic and cationic SIP highlighted their potential for controlled kinetics (chain length) and copolymerizability.[37]
Our early work on living anionic surface-initiated polymerization (LASIP) showed this using flat surfaces and silica nanoparticles.[38] Flat silicon (SiOx), gold surfaces, and deposited clay surfaces were ideal substrates to demonstrate the growth of polymers, surface density, and copolymerizations.[39,40] Homopolymer and block copolymer brushes were then demonstrated on gold by LASIP in a polar solvent.[41] A type of surface-activated LASIP was also demonstrated using surface-bound 1,1-diphenylethylene (DPE) initiators.[42]

While most of the popular methods for polymer brush formation gravitated toward ATRP and RAFT due to the ease and versatility of the protocols, free radical initiation was investigated early and effectively, especially for thick polymer brushes.[43] To demonstrate the versatility of substrates, we reported early on the preparation of polystyrene brushes via free radical surface-initiated polymerization (FR-SIP) for solvent-dependent friction force response AFM,[44] and from clay nanoparticles adsorbed on a planar substrate.[45] A conjugated polymer network (CPN) is used to crosslink polyvinyl carbazole (PVK) films, and polymer grafting of hole-transport precursor polymer brushes on ITO electrodes is used to enhance device performance.[46] SI-ATRP was used to graft a stimuli-responsive polymer onto a nanolayered, coextruded, and patterned polystyrene/polycaprolactone (PS/PCL) film.[47]

Another advantage of SIP is its ability to copolymerize and exert topological control.[48] Statistical or random sequences follow copolymerization and reactivity ratios. However, living polymerization methods can enable the formation of block copolymer sequences. Another possibility is to control topologies resulting in microstructures from branched to cyclic or loop topologies.[49] Branching or hyperbranching is interesting but is hardly controllable for confined surface initiators, even with the use of inhibitors. The ability to control polymer brush density, microstructures, composition, and gradient properties is important for fields like lubrication and anti-fouling surfaces in the biomedical field.[13]

"Grafting onto" method is another method to fabricate polymer brushes.[50] Instead of using SIP (or "grafting from" approach) to produce high-density grafted polymer brushes, "grafting onto" methods can be important for applications where chain architectures can be linear, branched, amphiphilic, dendritic, hyperbranched, or in a hybrid system. The synthesis of these adsorbing or chemisorbing macromolecules and amphiphiles can proceed via macromonomers, macroinitiators, or block copolymerization.[51–55] For example, dual-mode patterning precursors are synthesized on a conducting surface via electrochemically deposited benzophenone moieties, which can be photo-activated for polymer brush grafting.[56]

Here are some other examples of our work related to the "grafting onto" route. We have investigated the grafting of polymer loops and brushes on surfaces. For example, we have focused on adsorption kinetics and viscoelastic behavior of α,ω-thiol telechelics on gold.[57] Another approach we have taken is the electrochemical method for grafting polymers onto surfaces. We first reported the designing and synthesis of these adsorbing macromolecules via LFRP or RAFT polymerization, yielding dendron-end-functionalized, electroactive polymers, and initially demonstrated their surface-grafting properties.[58] The anti-biofouling properties of dendron-anchored, electrografted oligoethylene (OEG) or OEGylated surfaces were reported, along with their resistance to protein adsorption.[59] The monolayer adsorption of thiol-terminated, dendritic oligothiophene macromolecules was also prepared on gold surfaces.[60] We also fabricated poly(9-vinylcarbazole) thin films by photo-grafting the polymers to chemically bound benzophenone groups (via SAM) on ITO, monitoring the kinetics and thickness of the deposited films.[61,62] These films were then effectively used as a polymer hole transport layer in OLED devices.[63] A unique method of grafting onto colloidally patterned surfaces is to employ azide click chemistry and electro-graft onto conducting polymer arrays with complementary units.[64] In addition, ring-opening metathesis polymerization (ROMP) is an interesting living polymerization method that should be versatile for producing polymer brushes.[65,66] We demonstrate nanoparticle formation along with ultrathin film electrodeposition of carbazole-dendronized polynorbornenes, via ROMP.[67]

In general, polymer brushes are a strategy involving ultrathin polymer films and hybrid films. This includes multilayer nanostructures derived from monolayers, self-assembled monolayers, chemisorption, nanoparticles, Langmuir-Blodgett (LB) films, sol-gel films, and other hybrid layering methods.[68–71]

The ability to fabricate nanostructured ultrathin films includes layer-by-layer deposition of polyelectrolyte multilayers ( LBL-PEM) and other self-assembly or polymerization processes. They have been investigated by many groups, including us.[72] Opposite-charge complexation, exchange, or ion displacement were driving forces for film formation. We have combined surface-initiated polymerization (SIP) with LBL and macroinitiators to fabricate grafted polymer brushes.[73] Using this method, we fabricated free-standing films of semi-fluorinated block copolymer brushes from the LBL-PEM macroinitiators and demonstrated isolation as ultrathin bilayer films.[74] A binary mixed, stimuli-responsive polymer brush was grafted using the LBL-SIP method and fabricated likewise as free-standing films.[75] With varying surface initiator density, a Langmuir−Schaefer (LS) macroinitiator was deposited on a substrate for grafting a thermosensitive polymer brush via surface initiated- atom transfer radical polymerization (ATRP).[76] We also demonstrated dual-stimuli-responsive polymer films from a binary architecture via a combination of an inner PEM layer and SIP brush approach, achieving distinct pH- and temperature-responsive properties.[77]

In addition, it is possible to use crosslink chemistries to achieve strong hydrogel and viscoelastic properties.[78] This is mostly done by photochemical or chemical crosslinking strategies. Surface hydrogels are capable of controlled release.[79] They can also be classified as polymer brushes and polymer bottle brushes similar to highly and densely tethered brush systems except with a single chain or 1-dimensional polymer chain substrate.[80] For example, surface-grafted rewritable brushes with dynamic covalent linkages were demonstrated using pH- and thermo- responsive poly[2-(dimethylamino)ethyl methacrylate] (PDMAEMA).[81] We have also pioneered a number of methods on electrochemical crosslinking in ultrathin films with a combination of layer-by-layer polyelectrolytes, precursor polymer networks, and electropolymerizable side-group monomer precursors.[82]

Another interesting consequence is the production of multi-combinations with mixed brush compositions, gradient brushes, and mixtures of high and low brush lengths.[83] These multi-modal combinations can be done based on arrangements of mechanisms and latent chemistry. Since it is possible to control the activation of the initiators, it is possible to produce a mixture of initiators with a gradient of activations. They have been effective for capture-release mechanisms and sensing.[84] Polyelectrolyte brushes with polycationic, polyanionic, and polyzwitterionic systems are particularly interesting for ion transport and electrochemical behavior, including antibiofouling applications.[9,85]

An important development in patterning is the introduction of lithographic and non-lithographic methods.[86] This includes substrate patterns, photopatterning, chemical etching, and grafting-on methods.[87] We contributed significantly to colloidal nanosphere lithography (CNL), developing patterns and arrays from hexagonally packed colloidal submicron spheres (nanospheres) using LB-like deposition methods. This resulted in regular arrays of hexagonally packed structures that can be used for chemistry or polymer brush grafting. This is based on electropolymerizing monomers of conducting polymers such as polythiophene, polycarbazole, or polypyrrole, or functionalized monomers that, upon polymerization, become macroinitiators. An early review summarized our work on functionalizing and forming polymer brushes via surface-initiated living radical polymerization, forming nanopatterned polymer brush surfaces, or conducting polymers.[88] A more recent review emphasizes functional and stimuli-responsive polymers on micro- and nano-patterned interfaces.[89] One of our first reports revealed patterned surfaces combining polymer brushes and conducting polymer (polythiophene) via this colloidal template electropolymerization.[90] These colloidally templated two-dimensional conducting polymer arrays and SAMs result in binary compositional patterns and chemical functionality.[91] More recently, a macromolecular surface with binary brushes and conducting polymer patterns was shown to be capable of dual heat and light responses, via PET-RAFT.[92] We have demonstrated other unique methods, including templating of a lotus leaf with micro molding with PDMS and producing a PEM layer of ATRP initiators, resulting in a reversible superhydrophilicity and superhydrophobicity on a lotus-leaf pattern with poly-N-isopropylacrylamide (polyNIPAM) based thermo-responsive polymer brush.[93] Another involved reactive-ion etching for PEM patterning on Al "hard layers" and transfer printing to result in free-standing, self-folding polymer–metal bilayer particles, as well as subsequent growth of polymer brushes on the PEM layers with macroinitiators, as in the previous study.[94] In the area of electro-nanopatterning using conducting AFM, we introduced grafted polythiophene pendant polymer brushes and their electro-nanopatterning based on cantilever tip movement and voltage to close the tip-substrate circuit.[95]

A distinct polymer brush "grafting from-" and "grafting onto-" method developed by our group used electrochemistry to either form layers of electropolymerized macroinitiators or attach end-functionalized electropolymerizable groups to preformed polymer chains.[96,97] Another electrochemical approach is a "grafting-through" method for surface-anchored polymers. We demonstrated this by electrodeposition of an electroactive methacrylate monomer, which was then stitched via RAFT polymerization to form brushes.[98] The variations of these methods are as follows: An electrochemically crosslinked, surface-grafted polyvinyl carbazole (PVK) brush with an AIBN initiator was used as the hole-transport layer (on ITO) for organic photovoltaics. It showed a similar junction behavior to the PEDOT:PSS but without the acid instability.[99] An electrochemically active, dendritic−linear block copolymer via RAFT polymerization was synthesized and then characterized to form PS brushes via "grafting onto-" approach.[100] We first successfully reported grafting of polymers from electrodeposited macro-RAFT initiators on conducting surfaces using a "grafting from-" approach, and the electrodeposition was studied by cyclic voltammetry and AFM.[101] In a "grafting from-" approach, electropolymerized macroinitiators were used to graft various polymers that have superhydrophobic, superoleophilic, and hemi-wicking properties.[102] The electropatterning of binary polymer brushes (by SI- RAFT and SI-ATRP chemistries) was demonstrated on ITO electrodes.[103] In addition, we demonstrated the versatility in patterned polymer brushes through electrodeposited ATRP, ROMP, and RAFT initiators by applying colloidally templated arrays.[104] A surface-Initiated, ring-opening metathesis polymerization (SI-ROMP) was deployed to create polymer brushes, in which electropolymerizing terthiophene functionalized, olefin peripheral dendrons enabled ROMP to achieve high-density polynorbornene brushes with a Type II Grubbs catalyst.[105] A patent was issued for methods on preparing polymer coatings via electrochemical grafting of polymer brushes and related compositions.[106]

Photo-electron transfer-RAFT (PET-RAFT) is an important technique for carrying out RAFT polymerization under ambient conditions.[107,108] We reported in a recent review on PET-RAFT polymerization under flow chemistry and on surface-initiated reactions to create polymer brushes and patterned films.[109] We reviewed on SI-PET-RAFT polymerization by using electrodeposited macroinitiator thin films for biomedical and sensor applications.[110]

Hyperbranched polymers were prepared by combining PET-RAFT with a self-condensing vinyl polymerization (SCVP) under a flow chemistry setup.[111] Lastly, we reported the synthesis of hyperbranched polymer films by using electrodeposition and oxygen-tolerant, photoinduced SIP polymerizations.[112]

## 2.1 Interfaces, Colloidal Particles and Hybrid Grafting:

Polymers are synthesized in solution, on surfaces, in colloids, or in heterophase conditions. Therefore, polymerization conditions for polymer brush grafting can also be extended to various media and interfaces: solid-gas, solid-liquid, liquid-gas, and liquid-liquid.[113] Other possibilities include hybrid nanoparticle polymer brushes that incorporate nanoparticle transport and interface interactions. Polymer brushes are grown at the liquid-liquid interface, where the initiator, monomer, or chain-transfer agent is in a separate phase.[114] Polymers are also grown in the vapor phase, both on flat substrates, colloidal particles, or solid-gas interfaces.[115] Another possibility is to examine other thin-film assemblies, including layer-by-layer PEM substrates and LB film substrates.

Colloidal particles and nanoparticles are considered high surface-to-volume substrates for polymer grafting chemistry.[116] It is also a pathway to produce more polymer for ex-situ post-polymerization analysis. The particle and polymer layer is a core-shell particle geometry. This is very important for diverse applications, from plasmonic sensing, bulk powder handling, to the surface reactivity of nanocomposites. As a model substrate for SIP, it lends itself to several colloidal analytical characterization methods and has successfully demonstrated with layer-by-layer polyelectrolyte multilayer shells and hollow shells.[117]

## 2.2 Star Copolymers microstructures and Hybrids: Zero D polymer brushes

Star polymers are like zero-dimensional substrates for polymer brush.[118,119] At best, they can have 3D substrate properties with swelling behaviors.[120] We have utilized star copolymers and copolymers to demonstrate their potential for grafting to surfaces and, at the same time, stabilizing or mediating nanoparticle stability. Our first report using this approach involved preparing aggregation-stable gold nanoparticles via star-block copolymers prepared by anionic polymerization.[121] Star-like copolymer approach was also used to stabilize other noble-metal nanoparticles such as Pd and Pt.[122] A core–shell gold nanoparticle-star copolymer composite with gradient transfer and transport properties was reported in view of harnessing them for electro-optical sensors and catalysis applications.[123] A guest-host system approach was taken to form macromolecular particles based on a star copolymer with electrochemically active peripheral carbazole units.[124] These star copolymers are useful for novel, scalable controlled-release systems incorporated into polymer nanosheet layers.[125] The compilation of methods to prepare particles and nanoparticles by grafting initiators to enable SIP is a common theme in polymer brushes. We had previously reviewed these methods.[126]

We were among the first to demonstrate grafting of polymers from clay nanoparticles via in situ FR-SIP.[127] We distinguished SIP efficiency between monocationic and bicationic intercalated AIBN-type initiators.[128] Narrowly polydispersed PMMA by the FR-SIP methods was reported from these intercalated AIBN/clay nanoplatelets compared to solution-based reactions.[129] A comparison was also made between free radical vs. anionic initiator clays.[130] The degradation and relaxation kinetics of polystyrene (PS)–clay nanocomposite prepared showed a higher glass transition temperature (Tg) and high thermal stability.[131] Many other groups have reported the grafting of polymer brushes on chalcogenide nanoparticles, and some of the works involved functionalizing these nanoparticles with dendronized ligands.[132]

Graphene nanomaterials and hosts of SIP are of particular interest because of their electro-optical properties, but also rich chemistry for dispersions, colloidal systems, and emulsions.[133] Increasingly, graphene oxide (GO) or reduced GO is more commercially available and will likely be relevant for industrial applications. Some of our work in this area is as follows:

The functionalization and chemical modification of graphene oxide are key steps fpr their applications.[134] We reported on graphene oxide bifunctionalized with $NH_2/NH_3+$ as well as their outstanding corrosion resistance.[135] A simultaneous reduction and functionalization of GO via Ritter Reaction exhibited efficient conversion.[136] A high-volume, selectively monofacial modification of GO nanosheets was reported in suspensions with the aim of forming Janus nanoparticles.[137] Grafted carbazole-GO was used for electrodeposing ultrathin GO films on a variety of metal substrates.[138]

Many applications or potential applications were demonstrated. For example, grafted GO nanoparticles were used as a yield point enhancer in water-based drilling fluids.[139] GO–Poly(ethylene glycol) methyl ether methacrylate nanocomposite hydrogels showed high rheological stability.[140] We described on polymer-grafted graphene via ATRP, which exhibited enhanced rheology in oil-based drilling fluids.[141] Polymer-carbon-based nanomaterial filters were demonstrated for simultaneous removal of bacteria and heavy metals, using modified GO.[142] We reported on the use of aromatic-functionalized reduced graphene oxide to confine corrosive ions in nanocomposite coatings.[143] The Role of α, γ, and metastable polymorphs on electrospun polyamide 6/functionalized GO was described.[144] The optimization of the mechanical and setting properties of acrylic bone cements containing GO has been reported.[145]

In many high-performance applications using GO and modified GO, carbon fiber, and carbon nanotubes (CNT), the modification procedure is important, especially for high-performance composites and fiber composites. We have reported on the influence of surface modification on the performance of carbon fiber/epoxy composites.[146] Another example used crown ether-based supramolecular adhesives to enable durable adhesion in carbon fiber composites.[147]

Colloidal particles can also be coated with LBL-PEM shells and even prepared as hollow-shell capsules using perfect spheres such as polystyrene (PS) or polymethylmethacrylate (PMMA) as substrates.[148,149] LBL-PEM assemblies of polyelectrolytes can be extended from nanostructured films to colloidal nanoparticles via core-shell coating and hollow-shell formation.[117] We examined the LBL shell self-assembly of polyaniline (PANI) and sulfonated polystyrene (PSS), as well as the procedure to form multilayer-coated colloidal particles and hollow shells.[150] To further detect polyelectrolyte assembly, cross-linked, luminescent, spherical colloidal particles as well as hollow-shell particles were formed using polyfluorene-labeled polyelectrolytes. Grafted polymer brushes and a fuzzy ternary particle system were prepared by combining SI-ATRP and an LBL-colloidal core−shell system, with the outer layer being a macroinitiator shell.[151]

The potential for high-volume synthesis of polymer-grafted colloidal nanoparticle reactors can be demonstrated using continuous-flow chemistry (CFC) reactors. CFC reaction methods are used in organic synthesis.[152] We have also reported its use in polymer synthesis.[153] The continuous flow chemistry and fabrication of block copolymer–grafted silica micro-particles was carried out in an environmentally friendly water/ethanol medium.[154] Hyperbranched PDMAEMA-functionalized $SiO_2$ microparticles were demonstrated using ATRP polymerization and grafting in a flow reactor.[155] It should be noted that the method of grafting from particles can also be demonstrated with large sand particles, as in the case of polymerized hydrogel shells on sand used as suspending proppants for hydraulic fracturing.[156]

The polymer brush behaviors at solid-gas interfaces are an urgent topic.[157] Humidity control with other gases can have enormous potentials.[158] For instance, antimicrobial graphene/rubber composite via SI-ATRP were reported,[159] which used a Surface-Initiated PhotoATRP (SI-photoATRP) and red-light irradiation for open-air brush growth via multivariate experiments. The ATRP reaction is based on MB+-Cu/ L-catalyzed activation on various different surfaces.[160]

Vapor deposition methods enable monomers to be evaporated and deposited on surfaces and substrates with initiators or reactive groups.[161] Physical vapor deposition (PVD) can be used to evaporate monomers and initiators, with a specific bias introduced by an electric field or ion-assisted deposition. Some of these examples are highlighted as follows:

SAM modifications on ITO surfaces was studied during SI-vapor-deposition-enabled polymerizations of carbazole films.[162] We demonstrated the interfacial control and subsequent applications in enhanced organic light-emitting devices (OLEDs).[163] The carbazole polymer thin films were fabricated by chemically attaching to the substrates via PVD and SAM.[164] A high-vacuum vapor deposition of N-carboxy anhydride (NCA) benzyl glutamate polymerization was reported, along with *in situ* monitoring.[165]

## 2.3 In Situ and Ex Situ Characterization of Polymer Brushes and Hyphenated Methods

If the high interest and breadth of the polymer brush chemistry, microstructure, nanostructuring, and patterning are not enough, then the interest in their characterization among materials scientists and physicists is equally voluminous. In general, this can be divided into two categories: thin-film characterization methods and colloidal particle characterization methods.[166] These are based on analytical techniques such as spectroscopy, microscopy, optical methods, scattering, electrochemistry, and dielectric properties. The examples are as follows: Starting with ultrathin films and solid-state confinement, techniques such as FT-IR-ATR, XPS, AFM, SPR, ellipsometry, EIS, QCM, SIMS, X-ray reflectometry, and neutron reflectometry can be used.

With colloidal and nanoparticle substrates, light scattering, optical microscopy, fluorescence microscopy, TEM, SEM, X-ray, and neutron scattering, etc. can be applied. Regarding thickness measurements, dielectric properties, polymerization kinetics, diffusion kinetics, ion /electron transport behavior, and tribology, it is of great interest to apply ex-situ, in-situ, real-time, or hyphenated experiments and measurements on these systems. Over the years, our group and collaborators have focused on innovative and improved analytical techniques for polymer brushes. Some of these works are highlighted below.

Single-molecule spectroscopy was utilized to investigate heterogeneous transport of molecular ions in polyelectrolyte brushes, suggesting diffusion characteristics of charged brushes dependent on grafting density as well as interaction potentials between probe ions and polymer chains (with solvent, pH, and concentration dependencies).[167] Fluorescence correlation spectroscopy (FCS) was applied to explore single-ion diffusive transport behaviors in Poly(styrene sulfonate) polymer brush matrices, revealing that coulombic interactions are more impactful than brush density changes.[168] The pH and charge effect on polymer-brush -mediated transport was examined by conventional and scanning fluorescence correlation spectroscopies.[169] We also studied the permeability of anti-fouling PEGylated polymer brushes by FCS.[170]

Sum frequency generation (SFG) was used to probe polymer-solvent interaction and conformational changes at molecular levels.[171] The findings indicated the significance of solvent-mediated deformation and aggregation on polymer surfaces, which can be applicable to more sophisticated polymer brushes.

Surface plasmon resonance (SPR) and plasmonic materials play an important role in polymer brush characterization because of the localization of SPR (LSPR) and the evanescent wave in ATR experiments that measure precise thickness and dielectric properties, especially when coupled with goniometry and waveguiding. We summarized our efforts and contributions to this field in a previous review article.[172]

Electrochemistry, when paired with SPR spectroscopy, is also a powerful tool for probing polymer thin films and polymer brushes.[173] We have also used SPR spectroscopy for investigating molecularly imprinted polymer (MIP) sensors.[174] Anti-biofouling properties are of high importance, and we have used SPR to investigate interfacial behaviors of oligoethylene glycol /linear dendron monolayers, with an emphasis on their aggregation and electropolymerizability.[175] These electropolymerized, bioresistant coatings showed that dendron design parameters influenced their protein resistance, as measured by SPR.[176] Most of our SPR analytical work is embedded with the numerous polymer brush systems we have reported through the years.

Quartz crystal microbalance QCM and dissipation QCM (QCMD are some of the most readily accessible probes for ultrathin polymer films, and together with SPR, complement the picture of polymer properties.[177–179] They have also been used for polymer brushes. Our early work involved probing amphiphilic polymer brushes by physisorption, such as the adsorption behaviors of Polystyrene−Polyisoprene diblocks with zwitterionic groups, along with the effects of varied microstructures and volume compositions.[180] We also quantified the adsorption and kinetics of multi-zwitterionic end-functionalized polymers with gold surfaces.[181] In the future, machine intelligence-centered systems can be used for automated characterization of these materials and interfaces, with ML-driven workflows and agentic AI with LLMs.[182]

Lastly, there are several possibilities for combining fast optical FT-IR imaging and spectroscopy based on focal plane array (FPA) technology in polymer films. The spectrochemical methods can reveal morphological, pattern, and domain information, along with chemical functional group information.[183] We have applied this method in our analytical work of polymer brushes and patterns, integrating it with the polymer brush systems we have reported through the years.

There are plenty of opportunities to integrate chemometrics, data, and data analytics with modeling and high-throughput experimentation. LLMs and agentic AI tasks can convert these powerful analytical and characterization methods into more ML-driven approaches for optimization and rapid analysis. Robotics and automation will enable higher throughput for sampling or automated experimental tasks. The creation of simulated data or images and integration with deep learning will upgrade digitally raw images or enable faster data analysis of spectra.

# 3. Simulation of Polymer Brushes

Polymer brushes are dynamic systems.(**Figure 2**) However, theoretical models begin with static properties in the solid state due to the high density.[184–189] In condensed phases, non-covalent interactions resemble the transition from a semi-crystalline to a glassy state. The surface grafting density is much smaller than the polymer radius of gyration (Rg) in solution. An ideal model is a monodisperse macromolecule grafted to a flat, planar, cylindrical, or perfect spherical substrate.[184–189] But this is hardly the reality when coupled with the chemistry, which does not guarantee a 100% yield in activation or equal transport and diffusion behavior with the growing chain, as in SIP. The initial modeling methods employed are the blob concept, self-consistent field theories (SCFT), and scaling relationships. This can account for N, the number of monomers, *p,* is the characteristic segment length of the polymer, grafting density (σ), thickness (h), surface area (A), and swelling parameters. σ = N$p^2$/A, *p* and *h* ∼ Nσ$^{1/3}$.[184]

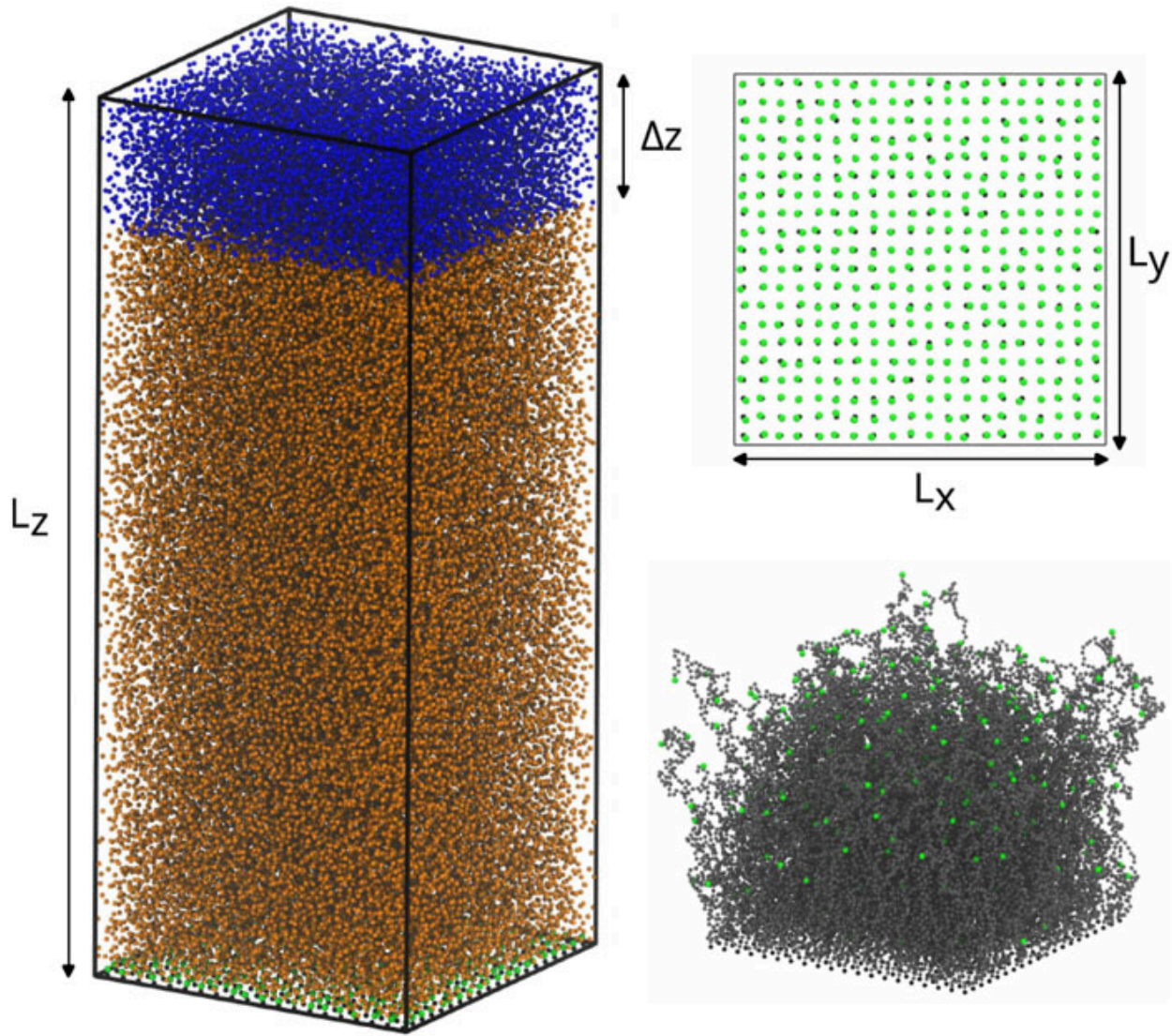

**Figure 2**. Simulation snapshot of the initial configuration (left), distribution of initiators on the substrate (top right), and grown polymer brush (bottom right). Grafted monomers, active sites (and initiators), free monomers, and bonded monomers are colored black, green, orange, and gray, respectively. The simulation box has dimensions 51.64 $\sigma$ × 51.64 $\sigma$ × 130 $\sigma$. Monomers are exchanged from the top of the simulation box of width $\Delta z$ = 20 $\sigma$ indicated in blue. The system shown is for $\varsigma$ = 0.150 $\sigma^{-2}$, and the brush conformation is for $\langle N \rangle$ = 80.16. (Figure and caption from reference[190] with no change under a creative commons 4.0 license [191])

These include the classical approximation method, the lattice formulation, and its repetition (or limitation thereof), and bead-spring models, which can be represented as a continuum.[184–189] Computer MD simulations test the validity of these approximations. By employing DFT and single-chain mean-field theory (SCMFT), atomistic and interaction potentials can be used to refine MD and other ab initio Monte Carlo simulations. Valid experimental results can elevate the status of the theory and yield additional hypotheses to be tested, e.g. classical swelling regimes of brush, mushroom, and flat. This is important to investigate various conditions of polymer brush swelling, shear stresses, chain configuration and conformation, brush-to-brush penetration, and stimuli-responsive behaviors.[189] Assumptions about the relaxation behavior, as described by Rouse dynamics, for brush coatings are overly simplified. What is needed are new correlation functions for multi-orientation and components that account for grafting sites, center-of-mass positions, and end-group orientations. Similar properties, as described in the solution behavior, do not exhibit the same relaxation or exponential decay.[185]

Tribology and simulation are interesting in the context of the application of shear forces (friction). This can be done with nonequilibrium MD to model the results from AFM and lateral force microscopy.[188] An important paper comparing simulated and synthesized polymer brush profiles with their characterization results from ellipsometry, and neutron reflectometry was published recently.[190]

Other specific examples of simulation protocols in polymer brushes include the following:
An all-atom, fully atomistic description with MD simulations was established for grafted Poly(N, N-dimethylaminoethyl methacrylate) polymer brushes, which functions as a model for polyelectrolyte brushes.[192] To understand the antifouling zwitterionic polymer brushes, a model was constructed with combined molecular mechanics (MM), Monte Carlo, MD, and steered MD (SMD) simulations.[187]

Because of the wide array of polymer brushes and surface-grafting strategies, current computational simulations and experimental designs of polymers often yield uncorrelated structures and properties across different parameter sets (no universal approach).[186] With data-driven ML approaches for simulations, they can be expanded beyond conventional research classifications. This can be started with more statistical methods, such as regression analysis, Bayesian optimization, decision trees, and various supervised learning methods, drawing on existing methods and results. Chemical or macromolecular descriptors are important. An ML workflow should incorporate screening methods with spatial, compositional, and interaction spaces. This could help streamline the focus to a range of polymers, brush microstructures, compositions, and a gradient approach to their design and synthesis. The goal is to accelerate data-driven, AI/ML-directed simulation with more structured empirical data. It can be an end-to-end workflow for scientists and engineers. This will, in turn, augment precision and efficiency in various industrial and scientific applications.[193]

In summary, this simulation section should have provided an overview of how these protocols can be advanced with AI/ML workflows, i.e., for the design, synthesis, and characterization of brushes. Advances in computational methods for thermodynamic models and molecular simulations will be augmented with ML approaches. For polymer brushes, accelerating understanding of SCP relationships is important and should also leverage LLM and RAG methods.

# 4. Introduction to AI/ML

The integration of AI/ML into polymer science marks a shift from traditional Edisonian trial-and-error to data-driven discovery.[193] At the heart of this paradigm transition is the evolution of statistical learning, where various mathematical or algorithmic architectures—supervised, unsupervised, and semi-supervised learning—allow pattern extractions from high-dimensional macromolecular datasets. By upgrading classical statistics with AI/ML algorithms such as Random Forests, Support Vector Machines, and Deep Neural Networks, researchers can now predict polymer brush properties, such as grafting density and conformational height.[194] To navigate the dimensionality in grafted polymers, Bayesian Optimization (BO) has emerged as an important approach for the iterative refinement in polymer brush experiments and simulations.[195–197] Unlike grid-search methods, BO uses a surrogate model—often a Gaussian Process—to represent the objective function, with an acquisition function to balance unknown regimes exploration and known optima exploitation .

The frontier of AI-driven synthesis is defined by the distinct but complementary roles played by Generative AI[198] and Reinforcement Learning (RL)[199–203]. Generative models, such as Variational Autoencoders (VAEs) and Generative Adversarial Networks (GANs), are utilized to "invent" novel monomer sequences or architecture motifs that satisfy specific physical constraints. On the other hand, RL focuses on the sequential decision-making process of synthesis. RL agents learn to adjust reaction conditions in real-time by receiving "rewards" based on the proximity of the synthesized polymer to a target molecular weight or polydispersity.

Recent advancements in Large Language Models (LLMs) and agentic AI are providing the critical infrastructures for fully Autonomous Self-Driving Laboratories (SDLs).[204] LLMs are now being designed to interpret protocols, extract historical data from legacy literature, as well as connecting robotic hardwares and operations through natural language commands. Agentic AIs can facilitate high-throughput experimentation (HTE) via data-driven experimental design, characterization results analyzing, or feedback-calibrated procedure redesign, with or without human intervention.

Translation of chemical structures into a machine-understable format is fundamental to such digital workflows, often achieved through the Simplified Molecular Input Line Entry System (SMILES).[205] SMILES strings allow for the digital representation of complex monomers and initiators as standardized text strings, which serve as an important part of descriptor generation.

To ensure that AI models are interpretable, rather than being just a pattern-extracting "black box," SHAP (SHapley Additive exPlanations) values are increasingly employed.[1,193,206,207] SHAP values provide an effective approach to quantify the contribution of each molecular descriptor—such as hydrophobicity or chain length—to the model's prediction, offering a transparent perspective into why an algorithm prefers a specific molecular descriptor for a given application like drug delivery or microfluidics.

# 5. AI/ML in simulation, design and future synthesis

In polymer brushes, the number of parameters, chemistries, and substrate geometries makes it very hard to form a universal algorithm and (coarse-grained) simulation approach.[194]

The application of both atomistic and MD simulations can yield substantial amounts of data and costly computer time. ML approaches can provide better pathways and, at the same time, provide improvements and a virtuous cycle with more data to discover novel properties and applications. This can be done by the following steps:

a) use LLMs and start with reliable datasets with RAGs and agentic AI tasks for specificity
b) employ correct molecular descriptors and strings for rapid training,
c) use ML algorithms (traditional ML or DL) with statistical and Bayesian optimization methods
d) use validation methods and feedback loop mechanisms with RL
e) venture into Generative learning as needed with autoencoders and GANNs

ML models and simulations can predict equilibrium behavior, interfacial forces, optimized gradient or sequence compositions, ion distributions in polyelectrolytes, and friction coefficients for specific hypothesis questions and applications to more dynamic environments. In general, ML can be used to design surfaces, including bulk coatings and high-performance materials.[208]

Some examples are as follows:

ML approaches were used to explore hydrogen-bond-charged polyanionic brushes, displaying steric and charge screening impacts on various counterions, revealing roles of water-polymer interactions and integrating MD simulations, ML-driven algorithms, and empirical data in an end-to-end AI/ML workflow protocol.[209]

ML enabled the prediction of the hydrophilic/hydrophobic properties of polymer brushes, using SMILES and SHAP values as descriptors, validation from contact angle measurements, and data presentation in MolLogP values to enable high-throughput correlation with wetting and surface-energy behavior.[210]

The ion transport and solvation effect in polymer brushes were identified by combining ML and MD techniques, revealing two hydration states in individual functional groups of cationic polymer brushes, as well as the influence of a specific counterion and brush length.[211]

An ML decision tree algorithm with SHAP values was deployed to forecast polymer surface free energy and to model chemical structure–property relationships, using data constructed from 72 types of polymer brushes in literature sources.[212]

An innovative study leverages ML to accelerate the synthesis of antifungal amphiphilic polymers by training a Random Forest binary classification model on curated data against fungal pathogens. It also detailed the early design parametrization with SHAP values.[213]

ML-optimized coarse-grained approaches were implemented to design polymer brush networks using ANN and BO methods, serving as a guide for future polymer composition and the corresponding thermo-mechanical properties.[214]

Remarkably, some of the most widely used models for ML-driven workflows of polymer brushes focused on improving anti-fouling and protein adsorption properties.[215] For example, regression and decision tree type ML methods were used to assess the impact of molecular weight (MW) and density on the of proteins adsorption of zwitterionic polymers in order to achieve optimal anti-fouling properties.[216] Random forests algorithms were developed to forecast the adsorption of serum proteins on polymer brushes, validated with data from five monomers and various processing parameters, which provided an important basis for refinement as well as a path for continuous simulation-experiment feedback cycles.[217] ANN and support vector regression (SVR) models held the promise of achieving superior polymer brush resistance to human blood serum adsorption.[218] Another example involved an ML approach to evaluate protein adsorption of a nonionic polyethylene glycol brush (compared to that of a zwitterionic polymer brush), in which both linear and nonlinear regression algorithms were used to evaluate grafting density, MW, brush length, buffer conditions, and temperature to improve correlations.[219] ML methods were also implemented to estimate and validate protein adsorption of bovine serum albumin and lysozyme on polymer brushes, with contact angle and the zeta potential as descriptors.[220] A data-driven ML with factor analysis of functional groups (FAFG), Pearson correlation analysis, ANN, RF, and Bayesian statistics are adopted to analyze structural chemical/surface features correlated with the antifouling behavior of self-assembled monolayers, applicable to polymer brushes.[221]

# 6. AI/ML in characterization and SDL

There are many reviews that outline analytical and characterization methods for polymer brushes in ultrathin films, colloidal particles, and at interfaces. In general, many of these relate to the coatings field (bulk or thin coatings) and self-assemblies or nanostructured films.[222] Many references and reviews can also be gathered in those systems regarding AI/ML workflows. It is also important to continuously find end-to-end examples of how simulation, synthesis, and characterization are augmented by this workflow. This section specifically focuses on how AI/ML workflows can augment the field of polymer brushes. Some examples are as follows:

## 6.1 Atomic Force Microscopy (AFM)

AFM is a powerful tool for investigating polymer brushes that combines many important variations in specific surface-probe microscopy (SPM) methods, including electrostatic, conducting, etc.[223] A previous review discusses approaches for improving AFM devices and data analytics using ML procedures.[224] This also refers to the context of specific SPM methods applied to polymer brushes throughout the years. The improvements focus on aspects of the AFM scanning process, data or image analysis, and the creation of a virtual AFM.(Figure 3) The workflow automates the recognition and measurement of domain sizes in polymer blends using deep learning (DL) and CNNs. Discrete Fourier transform (DFT), or discrete cosine transform (DCT) techniques can be adopted along with variance statistics and k-means based clustering, to examine phase separation.[225] ML-driven Electrostatic force microscopy (EFM) is used to investigate nanoscale dielectric property and permittivity in complex interphase geometries.[226] LLM and agentic AI are tasked to aid in automated AFM as AI lab assistants (AILA) for high-throughput experimentation.[227]

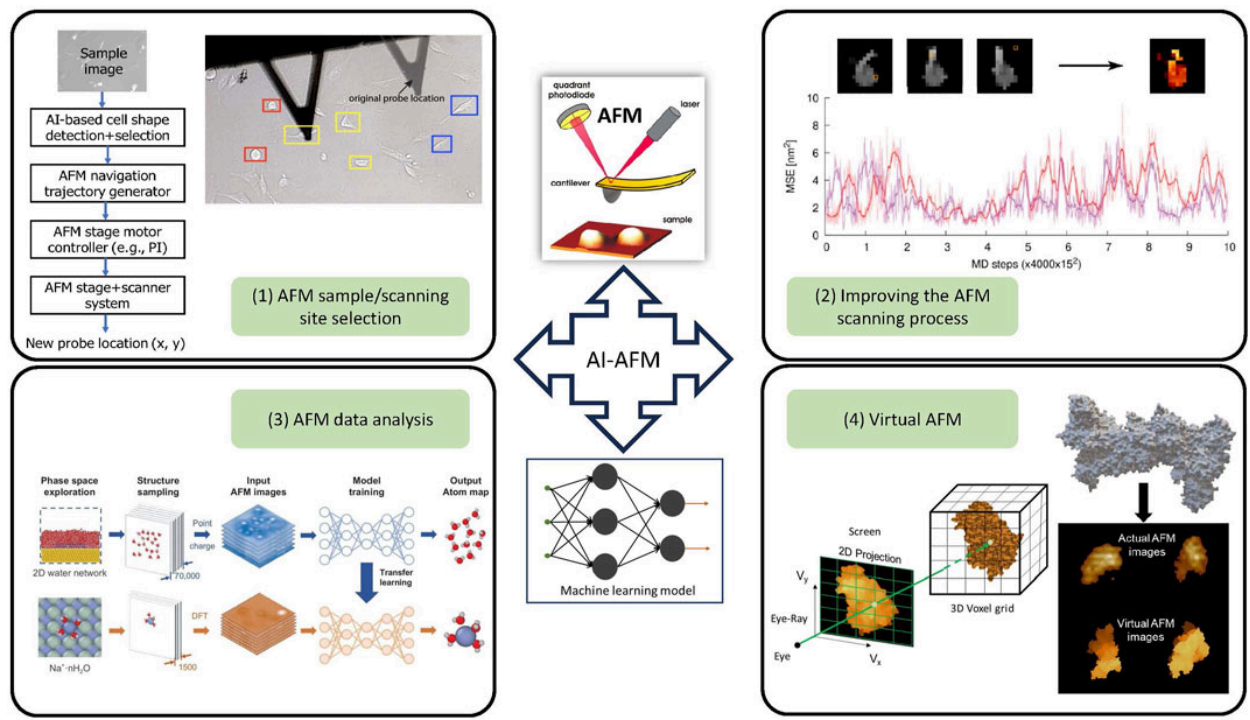

**Figure 3**. Overview of different steps of the AFM enhanced with the help of machine learning.[228] Machine learning models can improve both the experimental and the data analytics steps of AFM by improving (1) sample or scanning site selection ( developed ML-guided cell shape detection framework for automatic AFM tip navigation), (2) AFM scanning process ( developed a particle smoother (PS) method for Bayesian data assimilation to integrate molecular dynamics simulations with asynchronous HS-AFM movie data), (3) AFM data analytics (proposed using machine learning to identify atomic structures of interfacial water and ionic hydrates based on AFM images) and (4) creating virtual AFM ( proposed GPU-accelerated volume rendering technique to generate synthetic AFM images of protein samples). (Figure and caption from reference with slight rephrase under a Creative common 4.0 license [191] )

## 6.2 Surface Plasmon Spectroscopy (SPR)

Many applications of SPR are related to thin films, kinetics, biosensing, and therefore advances in polymer brushes or ultrathin films, as well as in adsorption phenomena.[229,230] SPR in polymer brushes has been used by many groups for *in-situ* and *ex-situ* characterization,[231] including us, especially in combination with electrochemistry. Recent reviews emphasized the use of ML methods for SPR and other plasmonic techniques in order to improve detection limits and achieve signal amplification.[232,233] For example, an ML-driven dip-searching method was demonstrated for angle-resolved SPR, which is valuable for modeling and automating a high-resolution analytical system for biosensing.[234] From our previous work on SPR and polymer brushes, and how we have utilized SPR-electrochemistry, it should be evident how these workflows will translate into future developments in the field.

## 6.3 Ellipsometry

Ellipsometry is an important and accessible method for investigating thickness, dielectric parameters in polymer brushes and their dynamic behavior.[235,236] The method, like other optical and dielectric methods, can be applied to both solid-state and metallic materials, as well as to dielectric materials such as polymers. The translation of these concepts to polymer brushes as dielectrics should be evident to those familiar with the instrumentation capabilities.

Applications of ML approaches are shown to be instrumental for works related to various ultrathin films, dielectrics, and metals, including polymer brushes, which are all suitable for existing ellipsometers and automated high-throughput optical characterization.[237] DL-based ellipsometry can enable rapid, high-accuracy assessment of optical constants and film components, with CNNs and transfer learning (TL) to enhance robustness.[238] A neural network powered algorithm was deployed to perform automated ellipsometry with excellent performance, validated by various materials (**Figure 4**).[239] Light-weight NN approaches were tested across a wide range of spectroscopic ellipsometry, accurately addressing inverse problems with only constrained computational resources.[240]

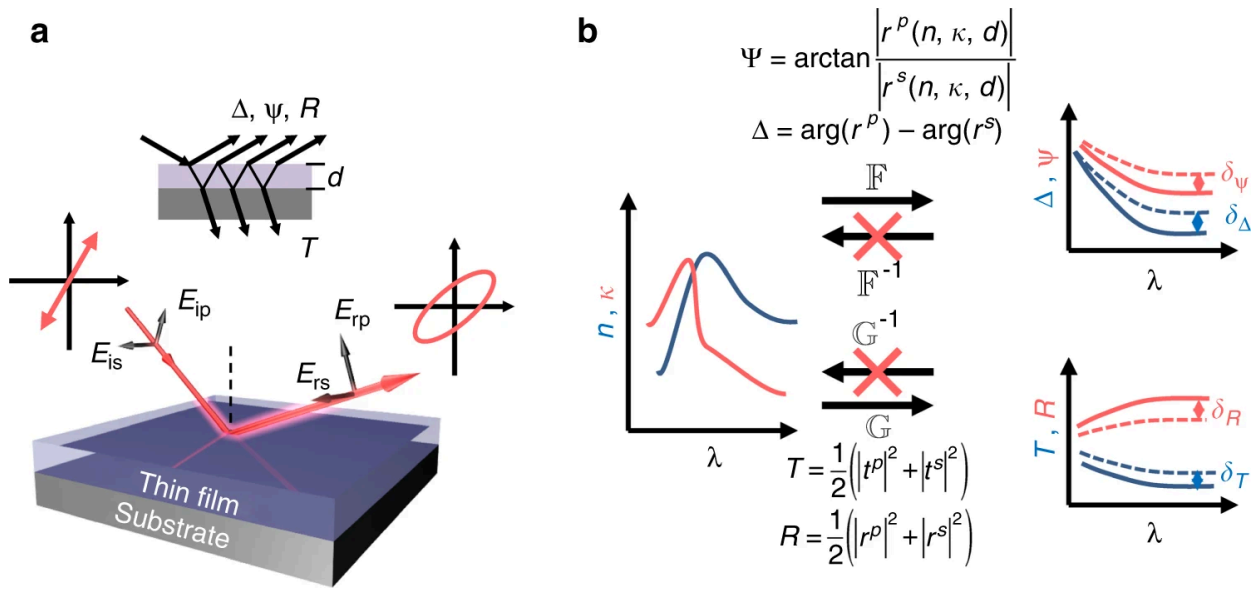

**Figure 4**. a In ellipsometry, light $E_i$ is obliquely incident onto a thin film with unknown ($n$, $\kappa$, $d$), which is on a substrate with known optical constants. Multiple beam interference occurs in the film, as shown in the inset. The orthogonal polarization components ($E_p$ and $E_s$) have unequal reflection coefficients $r_p$ and $r_s$, leading to elliptically polarized reflection $E_r$. $p$ and $s$ refer to polarizations with electric fields parallel and normal to the plane of incidence, respectively. b The ellipsometric angles $\Psi$ and $\Delta$ are defined by, $\tan\Psi\, e^{i\Delta} = r_p/r_s$, as denoted by . The total transmittance $T$ and reflectance $R$ are calculated by averaging the contributions from the $p$- and $s$-polarizations, denoted by . The physical process from ($n$, $\kappa$, $d$) to ($\Psi$, $\Delta$, $R$, $T$) has been well studied and can be modeled accurately as functions  and . On the other hand, the inverse, i.e., analytically inferring ($n$, $\kappa$, $d$) from ($\Psi$, $\Delta$, $R$, $T$), is generally impossible, and iterative fitting techniques have therefore been developed to numerically search a set of ($n$, $\kappa$, $d$) that best fits the experimental data. Solid lines denote experimentally measured ($\Psi$, $\Delta$, $R$, $T$), dashed lines present results recovered by the tentative ($n$, $\kappa$, $d$) solution (Figure and caption from reference[239] with no change under a  Creative common license 4.0 [191])

## 6.4 X-ray reflectometry (XRR) and Neutron Reflectometry (NR)

XRR and NR are powerful and efficient techniques for investigating structural and compositional characteristics of nanostructured and layered films.[241] Many examples of polymer brushes structural studies have been demonstrated with XRR and NR.[242] A previous study combined high-resolution, millisecond and monochromatic XRR with neural networks to reveal the kinetic growth of a polymer film.[243] A CNN model was deployed to showcase high-speed XRR for organic thin film grown at high vacuum deposition rates, which allows in-situ monitoring of thickness, density, roughness, and quantitative fits of XRR curves.[244] NN assessment of both NR and XRR allows predicting of different scattering length density (SLD) combinations for multi-component complex polymer systems.[245]

## 6.5 Quartz Crystal Microbalance (QCM)

QCM is an important and accessible method for analyzing both kinetics (in-situ and real-time) and static mass measurement of a thin film.[246] Its potential is dramatically augmented by AI/ML workflows.[178] Many examples have been reported for monitoring polymer brushes in situ and ex situ with QCM. QCM dissipation, or QCM-D, is a very informative, if not the method of choice, for analyzing polymer thin films and polymer brushes.[179] Combinations with impedance measurements also help refine the model with dielectric parameters supported with ML. ML was demonstrated in a QCM approach using regression and gradient boosting algorithms for sensing.[247] QCM results were analyzed utilizing a transfer-matrix framework that allows for long-range and asymmetric-gradient shear modulus, thus facilitating the identification of an glassy–rubbery interface in an immiscible polymer blend.[248] Furthermore, an ML strategy was applied to enhance QCM impedance measurement to increase the limit of detection (or LOD), which includes unsupervised k-means clustering along with supervised Support Vector Machine (or SVM) algorithm.[249]

## 6.6 X-ray Photoelectron Spectroscopy (XPS)

XPS is a potent elemental analysis tool that provides information on binding energies and oxidation states, with strong correlations among elements, especially with C, O, N, S, etc., which are most relevant to polymer materials and polymer brushes.[250] It is an essential tool for ultrathin film materials, both soft matter and solid-state, specifically for polymer brushes with multi-element compositions.[251] The use of ML algorithms for XPS analysis, complemented by SHAP, was established using ANN-based models to quantify element concentrations.[252]

Hybrid *ab initio* and reactive molecular dynamics (HAIR) and ML models were deployed to estimate the XPS elemental analysis at the Solid Electrolyte Interface (SLI) in lithium metal batteries.[253]

## 6.7 Fourier Transform Infra-Red (FTIR) Spectroscopy

FTIR is a spectroscopic technique for functional group identification in polymeric and organic thin films, with additional capabilities from polarization, photo-elastic modulation (PEM), grazing incidence, and specular reflection techniques.[254] It can be used in both *in situ* and *ex situ* settings. The techniques can be adopted for AI/ML workflows similar to the examples below: ML algorithm was utilized to carry out data clustering and automated qualitative analysis of FTIR results for graphene oxide dispersions.[255] FTIR of Poly (vinyl alcohol) /poly (vinyl pyrrolidone) blends were analyzed using linear discriminant analysis (LDA), k-nearest neighbors, and SVM.[256] DL was coupled with FTIR and Raman spectroscopy to perform microplastic analysis using one-dimensional CNN.[257]

## 6.8 Electrochemical Impedance Spectroscopy (EIS)

Impedance measurements are valuable for characterizing electron and ion transport mechanisms in polymer thin films.[258] EIS is an important technique to investigate the transport behavior of porosity in thin films. The method is highly sensitive and can be applied to various phenomena in thin-film dielectrics, including polymer brushes.[259]

ML algorithms in electrochemical analysis can increase sensitivity, selectivity, as well as dynamic range of diagnostic applications, which include EIS, voltammetry, amperometry, and photoelectrochemical sensing.[260] They are also considered label-free techniques. Some examples of augmentation with AI/ML workflows that can be adopted for polymers include the following:

AI/ML methods were used for corrosion analysis along with inhibitor design, adopting quantum, electrochemical, and image descriptors, as well as models such as SVR, RF, ANN, hybrid quantum ML model, and generative MoIGPT.[261] An ML model applied to polymer electrolyte brushes used an all-atom MD simulation and Linear Discriminant Analysis (LDA) algorithm for electro-osmotic (EOS) transport investigations in a cationic polymer-electrolyte -brush- grafted nanochannel.[262]

## 6.9 Autonomous SDL in the Synthesis and Characterization of Polymer Brushes

The classic design of an experiment (DOE) relies on a time-intensive methodology that is still labor-intensive even with experienced researchers.[263] The gap between fast and data-intensive simulation, LLMs, and high-throughput experimentation is significant. Clearly integrated, autonomous, and machine-intelligent experimental apparatus will enable *in-situ*, real-time methods for simultaneous dynamic characterization of optical, spectroscopic, electrical, gravimetric, and viscoelastic properties.[264] An autonomous self-driving lab (SDL) and material characterization can be incorporated with synthesis or deposition in a programmable, dynamic mode.[265] Data analysis and dynamic modeling, with real-time, iterative communication for dynamic control of experimental conditions, become feasible with a powerful edge computer. The goal is to introduce experts in the loop with minimal human assistance to enable time-efficient synthesis and characterization. This AI/ML-driven platform is envisioned for polymer thin-film and polymer-brush research to accelerate scientific discovery.[182]

In general, thin film fabrication and characterization for polymers and other dielectrics can be automated with the right mechatronics and robotic platforms (**Figure 5**).[266] Arraying, or the use of sample holders and programmable sampling bays, can accelerate the number of samples fabricated using a combination of DOE and analytical instruments. But a true SDL should involve more than this. It should be able to autonomously run experiments, apply LLMs and agentic AI-driven tasks to optimize results, and perform feedback loop iterations. This should complete an end-to-end AI/ML workflow for polymer thin films and polymer brushes.

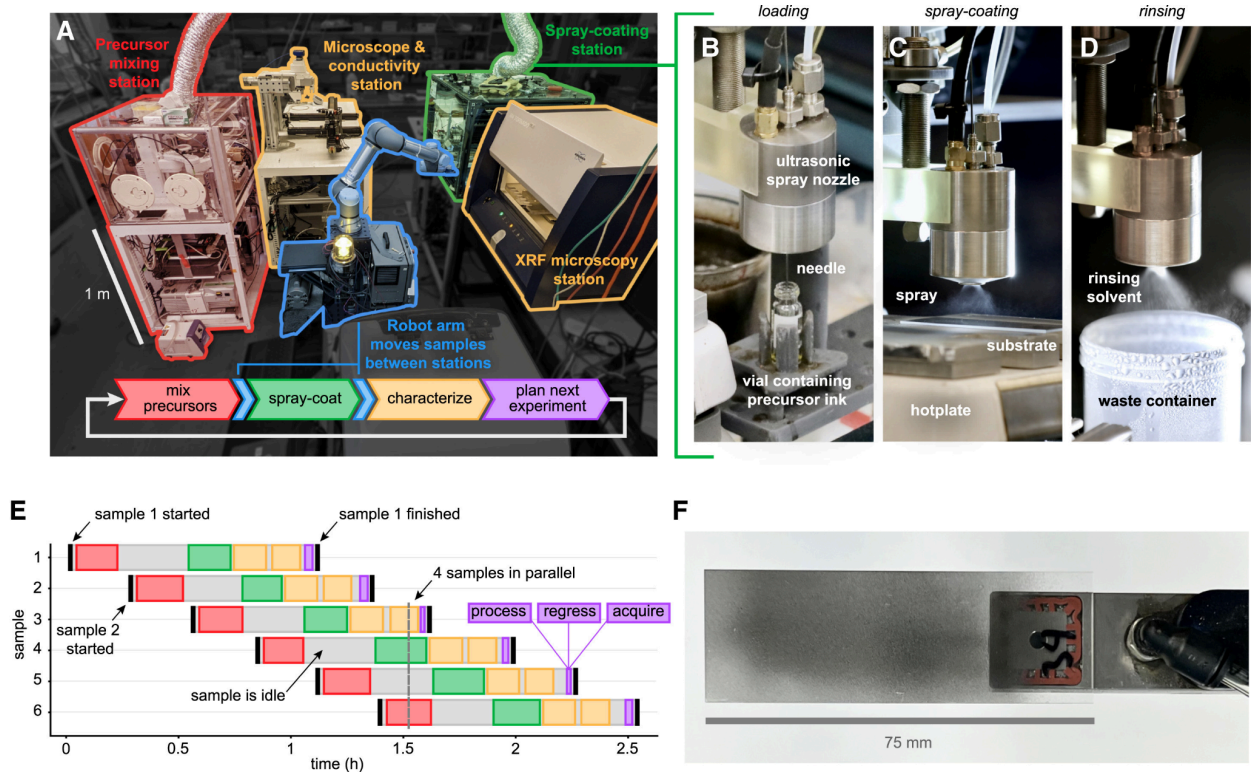

**Figure 5**. (A) The SDL consists of two synthesis and two characterization stations linked together by a central six-axis robotic arm. The robotic arm transports spray-coating precursor inks in 2-mL vials (B) and thin-film samples on 75 × 25-mm glass substrates (F) between stations as required to execute the parallelized closed-loop optimization workflow shown in simplified serial form in (A, inset) and in full detail in E). (B–D) The spray-coating station is custom designed to load precursor inks through a retractable needle that protrudes through the nozzle (B). Front loading minimizes the volume of precursor ink required to prime the nozzle. This enables small volumes (<1 mL) of precursor ink to be sprayed without wasting ink. The loaded precursor ink is sprayed on to a heated substrate to form a coating. (C). When spraying is complete, a solvent rinse is automatically performed to clear the nozzle of residual precursor ink (D). (E) A timeline of the parallelized workflow used for the optimization campaign. Ada's station architecture setup enables Ada to work on up to four samples in parallel. This parallel workflow enables around four experiments per hour and up to 100 spray-coating experiments per day. For clarity, the durations of the brief (<1 min) experiment-planning steps (purple) are not shown to scale. The experiment-planning process consists of three steps: data processing, GP regression, and acquisition of new experimental conditions. (F) The Pd film sample (left) is moved between stations by the six-axis robotic arm using a vacuum handler (right).
(Figure and caption from reference[267] with no change under a creative commons license BY-NC-ND 4.0 [268])

Some possible adaptations are also reported. A modular system for polymer analysis and discovery array, or PANDA, was reported.[269] This was designed to be a low-cost, self-driving lab for studying electrodeposited conducting polymers, such as the

poly(3,4-ethylenedioxythiophene): poly(styrene sulfonate) (PEDOT:PSS) film. In the PANDA, the electropolymerization variables, such as concentration, deposition duration, and voltage, were implemented via a BO method, leading to optimal electrochromic behavior. This SDL combines optical and electrochemical characterization, and therefore can be adopted to functional polymer brushes and films.

A comprehensive SDL in polymer electronics is demonstrated in Polybot, which streamlines synthesis and characterization of conducting polymers together with processing and fabrication of electronic devices.[270] An automated ML system for thorough polymer characterization and design addresses processing and nonequilibrium events, which enables hypothesis-driven projects.[271] An autonomous framework for solution polymer processing enables BO models towards low-defect, conductive polymer films.[272] Finally, a previous work outlines an open-source architecture for high-throughput analysis of organic thin-film transistor devices and organic films.[273]

# 7. Conclusions and Future Work

The transition of polymer brush research from the traditional "trial-and-error" methodology to autonomous Self-Driving Laboratories (SDLs) is an important ongoing paradigm shift in macromolecular science and engineering including polymer brush studies. By leveraging AI/ML-driven workflows, one can move far beyond simple data fitting type analysis into predictive synthesis and closed-loop optimization. This development allows rapid navigation of vast chemical, structural, and functional spaces, via effective exploration of the complex relationships between grafting chemistry, chain conformation, and macroscopic functionality. Future frontiers are expected in the following directions.

a. Transitioning to autonomous systems requires solving the hardware or "plumbing" of chemistry. This involves perfecting micro-reactors and microfluidics to handle viscous polymer solutions and integrating inline sensors for real-time AI feedback.

b. Significant opportunities lie in hybrid systems, such as AI-optimized "hairy" nanoparticles for targeted drug delivery and responsive brushes for high-sensitivity lab-on-chip diagnostics.

c. These AI workflows will not only redefine lab-scale formulations and coatings, but also the chemical, pharmaceutical, and biosensor industry. AI-designed polymer brushes will also hold promise in optimized bulk solid handling and create highly selective sorbents for the recovery of critical minerals.

The future of polymer brushes and macromolecular science lies in this closed-loop synergy, with AI navigating the chemical space and autonomous hardware executing the data-driven vision, transforming discovery from years-long processes into a previously unimaginably short amount of time.

## Funding and Acknowledgement

This work was supported by the Center for Nanophase Materials Sciences (CNMS), which is a US Department of Energy, Office of Science User Facility at Oak Ridge National Laboratory, and Laboratory Directed R&D (ORNL INTERSECT).

## Conflicts of Interest

The author declares no conflict of interest.

## Author Contributions

RCA planned the layout. RCA and JC all contributed to the overall design and writing.

(189) Desai, P. R.; Sinha, S.; Das, S. Compression of Polymer Brushes in the Weak Interpenetration Regime: Scaling Theory and Molecular Dynamics Simulations. *Soft Matter* **2017**, *13* (22), 4159–4166. https://doi.org/10.1039/C7SM00466D.

(190) Poudel, B.; Ritzert, P.; Robertson, H.; Soltwedel, O.; Humphreys, B.; Sodhi, M. K.; Kremer, K.; von Klitzing, R. Comparing Simulated and Synthesized Polymer Brush Profiles. *J. Chem. Phys.* **2025**, *163* (17), 174906. https://doi.org/10.1063/5.0287821.

(191) *Deed - Attribution 4.0 International - Creative Commons*. https://creativecommons.org/licenses/by/4.0/ (accessed 2025-03-19).

(192) Tippner, S.; Hernández-Castillo, D.; Schacher, F. H.; González, L. All-Atom Molecular Dynamics Simulations of Grafted Poly(N,N-Dimethylaminoethyl Methacrylate) Brushes. *J. Phys. Chem. B* **2025**, *129* (7), 2105–2114. https://doi.org/10.1021/acs.jpcb.4c07928.

(193) Cao, X.; Zhang, Y.; Sun, Z.; Yin, H.; Feng, Y. Machine Learning in Polymer Science: A New Lens for Physical and Chemical Exploration. *Prog. Mater. Sci.* **2026**, *156*, 101544. https://doi.org/10.1016/j.pmatsci.2025.101544.

(194) Ishraaq, R.; Das, S. Machine Learning for Probing Polymer and Polyelectrolyte Brushes. *Trends Chem.* **2025**, *0* (0). https://doi.org/10.1016/j.trechm.2025.10.011.

(195) Gongora, A. E.; Xu, B.; Perry, W.; Okoye, C.; Riley, P.; Reyes, K. G.; Morgan, E. F.; Brown, K. A. A Bayesian Experimental Autonomous Researcher for Mechanical Design. *Sci. Adv.* **2020**, *6* (15), eaaz1708. https://doi.org/10.1126/sciadv.aaz1708.

(196) Nambiar, A. M. K.; Breen, C. P.; Hart, T.; Kulesza, T.; Jamison, T. F.; Jensen, K. F. Bayesian Optimization of Computer-Proposed Multistep Synthetic Routes on an Automated Robotic Flow Platform. *ACS Cent. Sci.* **2022**, *8* (6), 825–836. https://doi.org/10.1021/acscentsci.2c00207.

(197) Biswas, A.; Liu, Y.; Creange, N.; Liu, Y.-C.; Jesse, S.; Yang, J.-C.; Kalinin, S. V.; Ziatdinov, M. A.; Vasudevan, R. K. A Dynamic Bayesian Optimized Active Recommender System for Curiosity-Driven Partially Human-in-the-Loop Automated Experiments. *Npj Comput. Mater.* **2024**, *10* (1), 1–12. https://doi.org/10.1038/s41524-023-01191-5.

(198) He, R.; Cao, J.; Tan, T. Generative Artificial Intelligence: A Historical Perspective. *Natl. Sci. Rev.* **2025**, *12* (5), nwaf050. https://doi.org/10.1093/nsr/nwaf050.

(199) Ali, Z.; Asparin, A.; Zhang, Y.; Mettee, H.; Taha, D.; Ha, Y.; Bhanot, D.; Sarwar, K.; Kiran, H.; Wu, S.; Wei, H. Automatic Design and Optimization of MRI-Based Neurochemical Sensors via Reinforcement Learning. *Discov. Nano* **2025**, *20* (1), 148. https://doi.org/10.1186/s11671-025-04338-z.

(200) Kim, S.; Cho, M.; Jung, S. Reinforcement Learning-Based Dynamic Optimization of Driving Waveforms for Inkjet Printing of Viscoelastic Fluids. *Langmuir* **2025**, *41* (17), 10831–10840. https://doi.org/10.1021/acs.langmuir.4c05141.

(201) Kiumarsi, B.; Vamvoudakis, K. G.; Modares, H.; Lewis, F. L. Optimal and Autonomous Control Using Reinforcement Learning: A Survey. *IEEE Trans. Neural Netw. Learn. Syst.* **2018**, *29* (6), 2042–2062. https://doi.org/10.1109/TNNLS.2017.2773458.

(202) Volk, A. A.; Epps, R. W.; Yonemoto, D. T.; Masters, B. S.; Castellano, F. N.; Reyes, K. G.; Abolhasani, M. AlphaFlow: Autonomous Discovery and Optimization of Multi-Step Chemistry Using a Self-Driven Fluidic Lab Guided by Reinforcement Learning. *Nat. Commun.* **2023**, *14* (1), 1403. https://doi.org/10.1038/s41467-023-37139-y.

(203) Su, Y.-T.; Lu, Y.; Chen, M.; Liu, A.-A. Deep Reinforcement Learning-Based Progressive Sequence Saliency Discovery Network for Mitosis Detection In Time-Lapse Phase-Contrast46

**Last update: 2.14.2026**